\documentclass[%
 reprint,
 amsmath,amssymb,
 aps,
prd,
nofootinbib
]{revtex4-1}

\newcommand\ee{\end{equation}}
\newcommand\be{\begin{equation}}
\newcommand\eea{\end{eqnarray}}
\newcommand\bea{\begin{eqnarray}}
\newcommand{\bV}{\mathbf{V}}

\newcommand{\bn}{\mathbf{n}}
\newcommand{\bx}{\mathbf{x}}

\newcommand{\bk}{\mathbf{k}}

\newcommand{\B}{\textrm{B}}
\newcommand{\F}{\textrm{F}}

\newcommand{\HH}{\mathcal{H}}
\newcommand{\LL}{\textrm{L}}

\newcommand{\Pdd}{P_{\delta\delta}}
\newcommand{\Pdv}{P_{\delta V}}
\newcommand{\Pdvdot}{P_{\delta\dot V}}
\newcommand{\Pdpsi}{P_{\delta\Psi}}
\newcommand{\Pvv}{P_{V V}}
\newcommand{\bB}{b_{\textrm{B}}}
\newcommand{\bF}{b_{\textrm{F}}}
\newcommand{\Eb}{\hat{E}^{\textrm{break}}}
\newcommand{\Opsi}{O^{\delta\Psi}}
\newcommand{\Ophipsi}{O^{\delta(\Phi+\Psi)}}
\newcommand{\Ostress}{O^{\rm stress}}
\newcommand{\aF}{\alpha_{\textrm{F}}}
\newcommand{\aB}{\alpha_{\textrm{B}}}

\usepackage{graphicx}
\usepackage{dcolumn}
\usepackage{bm}
\usepackage{amsmath,amsfonts,amssymb,slashed,upgreek,color}
\usepackage{multirow}
\usepackage{cancel}
\usepackage{hyperref}

\newcolumntype{C}[1]{>{\centering\arraybackslash}p{#1}}

\begin{document}


\title{Measuring anisotropic stress with relativistic effects}

\author{Daniel Sobral Blanco}
\author{Camille Bonvin}
\affiliation{D\'epartment de Physique Th\'eorique and Center for Astroparticle Physics,
Universit\'e de Gen\`eve, Quai E. Ansermet 24, CH-1211 Gen\`eve 4, Switzerland}


\date{\today}

\begin{abstract}
One of the main goal of large-scale structure surveys is to test the consistency of General Relativity at cosmological scales. In the $\Lambda$CDM model of cosmology, the relations between the fields describing the geometry and the content of our Universe are uniquely determined. In particular, the two gravitational potentials --that describe the spatial and temporal fluctuations in the geometry-- are equal. Whereas large classes of dark energy models preserve this equality, theories of modified gravity generally create a difference between the potentials, known as anisotropic stress. Even though measuring this anisotropic stress is one of the key goals of large-scale structure surveys, there are currently no methods able to measure it directly. Current methods all rely on measurements of galaxy peculiar velocities (through redshift-space distortions), from which the time component of the metric is inferred, assuming that dark matter follows geodesics. If this is not the case, all the proposed tests fail to measure the anisotropic stress. In this paper, we propose a novel test, $\Ostress$, which directly measures anisotropic stress, without relying on any assumption about the unknown dark matter. Our method uses relativistic effects in the galaxy number counts to provide a direct measurement of the time component of the metric. By comparing this with lensing observations our test provides a direct measurement of the anisotropic stress.
\end{abstract}

\pacs{Valid PACS appear here}
\maketitle

\section{Introduction}

Testing the law of gravity at cosmological scales is one of the main science driver for the coming generation of large-scale structure surveys. 
At large scale, the geometry of our Universe can be consistently described by two metric potentials, $\Phi$ and $\Psi$, describing perturbations around a homogeneous and isotropic background.\footnote{We use the metric convention $ds^2=a^2[-(1+2\Psi)d\tau^2+(1-2\Phi)d\bx^2]$ where $\tau$ denotes conformal time, and we neglect vector and tensor modes, that are negligible with respect to scalar modes in the linear regime.} Testing the law of gravity requires to test the relations between these two potentials and the energy-momentum tensor describing the content of our Universe, in particular the matter density fluctuation, $\delta$, and the galaxy peculiar velocity, $\bV$.
Two approaches can be used for this. The first one consists in assuming a specific model or class of models of gravity (e.g. Horndeski models~\cite{Horndeski1974}), determine how the four fields, $\Phi, \Psi, \delta$ and $\bV$, depend on the parameters of the model, and use observations (which depend on the four fields) to constrain the parameters. This approach has the obvious disadvantage that it has to be performed separately for each model or class of models.

The second approach consists in building model-independent tests, that allow to probe directly the relations between the four fields without assuming any model, see e.g.~\cite{Zhang:2007nk,Amendola:2012ky,Creminelli:2013nua,Kehagias:2013rpa,Motta:2013cwa,Ghosh:2018ijm,Bonvin:2018ckp,Franco:2019wbj,Bonvin:2020cxp}. The outcome of these tests can then be used to determine the validity of any theory of gravity. This second approach, which is more powerful, is however suffering from an important limitation: the fact that our observables at late time are sensitive to only three combinations of the four fields, namely $\delta$ and $\bV$ (through redshift-space distortions, see e.g.~\cite{Parkinson:2012vd,Alam:2016hwk}) and $\Phi+\Psi$ (through cosmic shear~\cite{Asgari:2020wuj,Gatti:2020eyt}, CMB lensing~\cite{Sherwin:2016tyf,Aghanim:2018oex,Wu:2019hek,Faundez:2019lmz} or Integrated Sachs Wolfe~\cite{Ade:2015dva}). This means that current observations are not able to test all relations between the four fields. The standard way of overcoming this problem is to assume that some of the relations between the four fields are known. Typically, one usually assumes that the continuity equation for dark matter holds: there is no exchange of energy between dark matter and dark energy; and that Euler equation for dark matter holds: there is no fifth force acting on dark matter, which consequently follows geodesics. Under these conditions, a measurement of $\bV$ can be translated into a measurement of $\Psi$, which can then be compared to $\Phi+\Psi$ to test if the two metric potentials are the same, i.e.\ to test for the presence of anisotropic stress~\cite{Amendola:2012ky,Amendola:2013qna,Pinho:2018unz,Arjona:2020kco}. This is a key test for modified theories of gravity since in General Relativity (GR) and for large classes of dark energy models, $\Phi=\Psi$ at late time~\footnote{Note that neutrinos also generate a non-zero anisotropic stress, which is however very small~\cite{Adamek:2016zes,Adamek:2017uiq}.}, whereas very generally in modified theories of gravity $\Phi\neq \Psi$, see e.g.~\cite{Saltas:2014dha}. 

In this paper we propose a novel model-independent test for the anisotropic stress, $\Ostress$, which does not rely on any assumption for dark matter, i.e.\ which does not rely on the validity of the continuity or Euler equation.  To build this test we use the fact that galaxy number counts are affected by gravitational redshift, a relativistic effect that is directly proportional to the field $\Psi$~\cite{Yoo:2009au,Bonvin:2011bg,Challinor:2011bk,Jeong:2011as}. We develop a method to isolate $\Psi$ from galaxy number counts observations. More precisely we build an observable which measures the correlations between the matter density fluctuations and the gravitational potential $\Psi$: $\Opsi\propto  \langle\delta \Psi\rangle$. We then compare this with lensing observations, which provide a measurement of $\Ophipsi\propto  \langle\delta (\Phi+\Psi)\rangle$. Our test is then simply given by the ratio between these two observables:
\be
\Ostress \equiv \frac{\Ophipsi}{\Opsi}=1+\eta\, ,
\ee
where $\eta$ relates the two metric potentials, $\Phi=\eta\Psi$. 
In $\Lambda$CDM, the two metric potentials are equal and therefore $\Ostress=2$ at all scales and redshifts. Any observed deviation from 2 would therefore unambiguously mean that the anisotropic stress is non-zero in our Universe, which is a strong indication for deviations from GR.

The rest of the paper is structured as follow: in Section~\ref{sec:method} we build the observables $\Opsi$ and $\Ophipsi$. In Section~\ref{sec:stress} we define our test, $\Ostress$. We then compare this test with the method presented in~\cite{Amendola:2012ky} to measure the anisotropic stress, which assumes the validity of the continuity and Euler equation for dark matter, and we show how this method breaks down if these assumptions are not valid. We conclude in Section~\ref{sec:conclusion}.

\section{Methodology}

\label{sec:method}

\subsection{Galaxy number counts}

Redshift surveys map the distribution of galaxies in the sky, providing a measurement of the galaxy number counts fluctuations
\be
\Delta\equiv\frac{N(\bn,z)-\bar N(z)}{\bar N(z)}\, ,
\ee
where $N$ denotes the number of galaxies per pixel detected in direction $\bn$ and at redshift $z$, and $\bar N$ is the average number of galaxies per pixel at redshift $z$. At linear order in perturbation theory, the dominant contributions to $\Delta$ are~\cite{Yoo:2009au,Bonvin:2011bg,Challinor:2011bk,Jeong:2011as} 
\begin{align}
\Delta(\bn, z)&=b\,\delta-\frac{1}{\HH}\partial_r(\bV\cdot\bn)\label{Delta}\\
&+(5s-2)\int_0^rdr'\frac{(r-r')r'}{2r}\Delta_\perp(\Phi+\Psi)(\bn,r')
\nonumber\\
&+\left(1-5s+\frac{5s-2}{r\HH}-\frac{\dot\HH}{\HH^2}+f^{\rm evol}\right)\bV\cdot\bn\nonumber\\
&+\frac{1}{\HH}\dot\bV\cdot\bn+\frac{1}{\HH}\partial_r\Psi\, ,\nonumber
\end{align}
where $\HH$ denotes the Hubble parameter in conformal time $\tau$, $r=r(z)$ is the comoving distance to redshift $z$, a dot denotes derivative with respect to conformal time and $\Delta_\perp$ is the Laplacian transverse to the photon direction $\bn$. The  functions $b(z)$, $s(z)$  and $f^{\rm evol}(z)$  are the  galaxy bias, the magnification bias and the evolution bias respectively. These functions depend on the population of galaxy which is observed as well as on the specifications of the survey.  

The first contribution in Eq.~\eqref{Delta}, $\delta$, is the matter density fluctuation in comoving gauge. The second term, which depends on the galaxy peculiar velocity, $\bV$, is the contribution from redshift-space distortion (RSD)~\cite{Kaiser:1987qv,Hamilton:1997zq}. The second line contains the effect of lensing magnification~\cite{Scranton:2005ci,Garcia-Fernandez:2016oud}. This contribution is subdominant with respect to density and RSD, except at high redshift~\cite{2014PhRvD..89h3535B,Jelic-Cizmek:2020pkh}. From~\cite{Bonvin:2020cxp} we expect this term to be negligible for our test, at least below $z=1.5$. In a forthcoming paper we will study this in more detail for specific surveys. The last 2 lines in Eq.~\eqref{Delta} contain the so-called relativistic effects, that depend on the galaxy peculiar velocity, through Doppler effects, and on the metric potential $\Psi$, through gravitational redshift. These relativistic effects have the specificity to generate odd multipoles in the power spectrum and correlation function~\cite{McDonald:2009ud,Yoo:2012se,2014PhRvD..89h3535B,Croft:2013taa}. As such they can be isolated from the dominant density and RSD contributions, which generate even multipoles~\footnote{Let us mention that the separation of $\Delta$ into RSD + relativistic effects is gauge-dependent. However the separation in odd and even multipoles is gauge-independent. In the flat-sky approximation, RSD contribute only to even multipoles whereas the relativistic effects in the third and fourth line of~\eqref{Delta} contribute only to odd multipoles. At large scales however, wide-angle and evolution effects mix the two types of contributions.}. In addition to these terms, $\Delta$ contains other relativistic effects that contribute to the even multipoles. These terms are however suppressed by $(\HH/k)^2$ with respect to density and RSD and can therefore be safely neglected.

Note that Eq.~\eqref{Delta} is valid at linear order in perturbation theory. In the non-linear regime, other terms have been shown to contribute to the observable $\Delta$, modifying the RSD contribution~\cite{Scoccimarro:1999ed, Nielsen:2016ldx}, and also the relativistic effects (like transverse Doppler effects~\cite{Zhao:2012gxk, Kaiser:2013ipa,Cai:2016ors, Beutler:2020evf}). The test proposed in this paper is valid only in the linear regime, where the impact from non-linear corrections can be neglected.

\subsection{Isolating gravitational redshift}

The aim of our work is to isolate the contribution from gravitational redshift given by the last term in Eq.~\eqref{Delta}, $\partial_r\Psi/\HH$, since it is directly proportional to the time component of the metric $\Psi$. The optimal way of targeting this contribution is to cross-correlate two populations of galaxies with different luminosities, such that this term contributes to odd multipoles~\cite{McDonald:2009ud,Yoo:2012se,2014PhRvD..89h3535B,Croft:2013taa}.
In Fourier space, the galaxy number counts fluctuations for a population of galaxies with luminosity $\LL$ becomes (we use the convention $f(\bk,\tau)=\int d^3\bx e^{i\bk\bx}f(\bx,\tau)$) 
\begin{align}
\Delta_\LL&(\bk, z)=b_\LL\delta(\bk,z)-\frac{k}{\HH}(\hat \bk\cdot \bn)^2 V(\bk,z)\label{Deltak}\\
&+i (\hat \bk\cdot \bn)\left[\alpha_\LL V(\bk,z)+\frac{1}{\HH}\dot{V}(\bk, z)-\frac{k}{\HH}\Psi(\bk, z)\right]\, ,\nonumber
\end{align}
\be
\hbox{where}\quad\alpha_\LL\equiv 1-5s_\LL+\frac{5s_\LL-2}{r\HH}-\frac{\dot\HH}{\HH^2}+f_\LL^{\rm evol}\, , 
\ee
and the velocity potential, $V$, is defined through $\bV(\bk,z)=i\hat\bk V(\bk, z)$. Eq.~\eqref{Deltak} is valid only in the flat-sky approximation, where $\bn$ can be considered as fixed. We will study in a future work the validity of this approximation for our test. 

The correlations between a bright, $\LL=\B$, and a faint, $\LL=\F$, population of galaxies are given by
\begin{align}
\langle \Delta_\B(\bk,z)\Delta_\F(\bk',z)\rangle =(2\pi)^3P_{\B\F}(k,\mu,z)\delta_D(\bk+\bk')\, ,\nonumber
\end{align}
where
\begin{align}
&P_{\B\F}=b_\B b_\F\Pdd-\frac{1}{3}(b_\B+b_\F)\frac{k}{\HH}\Pdv+\frac{1}{5}\left(\frac{k}{\HH}\right)^2\!\!\!\!\Pvv \nonumber\\
&+\left[-\frac{2}{3}(b_\B+b_\F)\frac{k}{\HH}\Pdv + \frac{4}{7} \left(\frac{k}{\HH}\right)^2\Pvv\right]L_2(\mu)\nonumber\\
&+\frac{8}{35}\left(\frac{k}{\HH}\right)^2\Pvv L_4(\mu)+\Bigg[ (b_\F\alpha_\B-b_\B\alpha_\F)\Pdv \label{PLM} \\
&+ (b_\F-b_\B)\frac{1}{\HH}\Pdvdot+ \frac{3}{5}(\alpha_\F-\alpha_\B)\frac{k}{\HH}\Pvv \Bigg] i L_1(\mu)\nonumber\\
&+\frac{2}{5}(\alpha_\F-\alpha_\B)\frac{k}{\HH}\Pvv \, iL_3(\mu)+ (b_\B-b_\F)\frac{k}{\HH}\Pdpsi\, i L_1(\mu)\, .\nonumber
\end{align}
Here $L_\ell$ denotes the Legendre polynomial of degree $\ell$, the angle $\mu=\hat\bk\cdot\bn$ and the power spectra are defined through
\be
\langle X(\bk, z) Y(\bk',z)\rangle=(2\pi)^3P_{XY}(k,z)\delta_D(\bk+\bk')\,,
\ee
for $X, Y = \delta, V, \dot V, \Psi$. 

Our aim is to isolate the last term in Eq.~\eqref{PLM}, which is proportional to $\Pdpsi$. The anisotropic stress can then be directly measured by dividing this contribution with the so-called galaxy-galaxy lensing correlation, which is proportional to $P_{\delta(\Phi+\Psi)}$~\cite{Prat:2017goa}. To isolate $\Pdpsi$, we first extract the dipole of $P_{\B \F}$ which is proportional to $\Pdpsi, \Pdv, \Pdvdot$ and $\Pvv$. We then look for combinations of the other multipoles in order to cancel the $\Pdv, \Pdvdot$ and $\Pvv$ contributions.

The multipole $\ell$ of $P_{\B\F}$ can be extracted by weighting it with the Legendre polynomial of degree $\ell$ and integrating over $\mu$
\be
P^{(\ell)}_{\B\F}(k,z)=\frac{2\ell+1}{2}\int_{-1}^1 d\mu L_\ell(\mu)P_{\B\F}(k,\mu,z)\, .
\ee

For our test, we need to measure the monopole and quadrupole of the bright and faint populations, the hexadecapole of the whole population and the dipole and octupole of the cross-correlation between bright and faint:
\begin{align}
P^{(0)}_\LL&=b_\LL^2\Pdd-\frac{2b_\LL}{3}\frac{k}{\HH}\Pdv+\frac{1}{5}\left(\frac{k}{\HH}\right)^2\Pvv\, ,\\
P^{(2)}_\LL&=-\frac{4b_\LL}{3}\frac{k}{\mathcal{\HH}}\Pdv + \frac{4}{7}\left(\frac{k}{\HH}\right)^2\Pvv\,,\\
P^{(4)}&= \frac{8}{35}\left(\frac{k}{\HH}\right)^2\Pvv\,,\label{P4}\\
P^{(1)}_{\B\F}&= -i\Bigg[(\bB\aF-\bF\aB)\Pdv + (\bB-\bF)\frac{1}{\HH}\Pdvdot\label{P1} \\
&\qquad\  -\frac{3}{5}(\aF-\aB)\frac{k}{\HH}\Pvv \Bigg]+ i(\bB-\bF)\frac{k}{\HH}\Pdpsi\,,\nonumber\\
P^{(3)}_{\B\F}&=i\frac{2}{5}(\aF-\aB)\frac{k}{\HH}\Pvv\, ,
\end{align}
with L=B, F.
From these observed multipoles we construct the following observables:
\begin{align}
O_\LL^{\delta\delta}(k,z)&\equiv P^{(0)}_\LL-\frac{1}{2}P^{(2)}_\LL+\frac{3}{8}P^{(4)}_\LL=b_\LL^2\Pdd(k,z)\, ,\\
O_\LL^{\delta V}(k,z)&\equiv \frac{3}{4}P^{(2)}_\LL - \frac{15}{8}P^{(4)}=-b_\LL\frac{k}{\HH}\Pdv(k,z)\, \label{odv} ,\\
O_\LL^{\delta\dot V}(k,z)&\equiv - (1+z)\sqrt{O_\LL^{\delta\delta}(k,z)}\frac{d}{dz}\left[\frac{O_\LL^{\delta V}(k,z)}{\sqrt{O_\LL^{\delta\delta}(k,z)}}\right] \nonumber\\
&= b_\LL\frac{k }{\HH} \left[\frac{\Dot{\HH}}{\HH^2}\Pdv(k,z) - \frac{1}{\HH}\Pdvdot(k,z)\right] \label{oddotv}\, .
\end{align}
To obtain the last line in Eq.~\eqref{oddotv} we use the fact that, in generic theories of gravity, the density and velocity fields in the linear regime can be written as $\delta(\bk,z) = D(k,z)\delta(\bk,0)$ and $V(\bk,z) = G(k,z)\delta(\bk,0)$, where $\delta(\bk,0)$ is a constant and denotes the present dark matter density, while $D(k,z)$ and $G(k,z)$ are functions of $k$ and $z$ mapping $\delta(\bk,0)$ into the past~\footnote{These relations are valid in theories of gravity where mode couplings are negligible in the linear regime.}. With this we can easily verify that
\begin{align}
\sqrt{\Pdd(k,z)}\frac{d}{d\tau}\left(\frac{\Pdv(k,z)}{\sqrt{\Pdd(k,z)}}\right)
&=D(k,z)\dot{G}(k,z)\Pdd(k,0)\nonumber\\
&=\Pdvdot(k,z)\, ,
\end{align}
which gives rise to expression~\eqref{oddotv}.
We are now able to isolate $\Pdpsi$ with the following combination
\begin{align}
O&^{\delta\Psi}(k,z)\equiv i\frac{\HH}{k}\left[\frac{3}{2}P^{(3)}_{\B\F}-P^{(1)}_{\B\F}\right] \nonumber\\
&\!\!-\left(\frac{\HH}{k}\right)^2\left[O_\B^{\delta\dot V}(k,z)-O_\F^{\delta\dot V}(k,z)\right] \nonumber\\
&\!\!-\left(\frac{\HH}{k}\right)^2\left[1-\frac{2}{r\HH}-5s_\F\left(1-\frac{1}{r\HH}\right)+f^{\rm evol}_\F\right]O_\B^{\delta V}(k,z)  \nonumber\\
&\!\!+\left(\frac{\HH}{k}\right)^2\left[1-\frac{2}{r\HH}-5s_\B\left(1-\frac{1}{r\HH}\right)+f^{\rm evol}_\B\right]O_\F^{\delta V}(k,z) \nonumber\\
&=(\bB-\bF)\Pdpsi(k,z)\,. \label{Opsi}
\end{align}

We see that $O^{\delta\Psi}(k,z)$ can be measured from the galaxy number counts, without making any assumption on the theory of gravity. It depends indeed on 
\begin{itemize}
\item The multipoles of the power spectrum, which are observable.
\item The background quantities $\HH/k$ and $r\HH$. These two combinations can be inferred from background observations. For example, observations of type Ia supernovae provide a measurement of the luminosity distance, up to a multiplicative constant~\footnote{This is due to the fact that the absolute intrinsic luminosity of supernovae is unknown, so that only ratios of luminosity distances at different redshifts are independent of normalisation.}: $\hat{d}_L\equiv d_L\HH_0$, from which one can infer the ratio ${\HH(z)}/{\HH_0}$. We then have
\begin{align}
r\HH&=\frac{\hat{d}_L}{1+z}\frac{\HH}{\HH_0}\, ,\quad\mbox{and}\quad\frac{\HH}{k}=\frac{\HH}{\HH_0}\frac{1}{\hat k}\ \, ,
\end{align}
where $\hat k\equiv k/\HH_0$ is independent of $h$ for $k$ in units Mpc$^{-1}h$.
\item The magnification bias, $s$, and evolution bias, $f^{\rm evol}$, of the bright and faint populations. These quantities can be directly measured from the two populations of galaxies. The magnification bias requires a measurement of the number of galaxies as a function of luminosity~\cite{Scranton:2005ci}, whereas the evolution bias requires a measurement of the number of galaxies as a function of redshift~\cite{Challinor:2011bk,DiDio:2013bqa}. 
\end{itemize}
The observable $\Opsi$ is, on its own, a very interesting quantity since it probes directly the correlations between density and gravitational potential $\Psi$. It provides therefore a way of measuring these correlations at cosmological scales, for the first time. 

This observable relies on the dipole and octupole of the power spectrum, that are too small to be measured with current surveys~\cite{Gaztanaga:2015jrs}. The signal-to-noise ratio (SNR) of the dipole has however been forecasted for the upcoming generation of surveys  (in configuration space), and is expected to reach 9.6 for the DESI survey (7.4 for the Bright Galaxy Sample and 6.2 for the emission line galaxies and luminous red galaxies~\cite{Bonvin:2015kuc}), and 46.4 for the SKA phase 2~\cite{Bonvin:2018ckp}. The octupole is between 2 to 5 times smaller than the dipole~\cite{2014PhRvD..89h3535B}, and its SNR is therefore expected to be reduced accordingly (its variance should indeed be similar, since it is dominated by density and RSD). The octupole may therefore degrade the overall SNR of $\Opsi$. If this is the case, this could be circumvented by replacing the octupole with the hexadecapole, which is similarly sensitive to $P_{VV}$, see Eq.~\eqref{P4}. Due to the relatively large SNR of the dipole for DESI and the SKA, we expect $\Opsi$ to be well measured in future surveys. We will study this in detail in a forthcoming work.

Let us mention that one limitation of the observable $\Opsi$ is to rely on the flat-sky approximation. Since relativistic effects may be of similar order as wide-angle effects, this approximation may not be accurate enough and wide-angle effects may modify the form of Eq.~\eqref{Opsi}. To study the importance of wide-angle effects, one needs to work in configuration space, where these effects can be consistently included~\cite{2014PhRvD..89h3535B}. In particular, in configuration space one can construct estimators that remove wide-angle effects directly from the signal~\cite{Hall:2016bmm,Bonvin:2018ckp}, without relying on any theoretical modelling. Adapting the observable $\Opsi$ to configuration space will not change its form, and we defer this to a future work.

\subsection{Galaxy-galaxy lensing}

To extract the anisotropic stress from $\Opsi$, we need in addition a measurement of $P_{\delta(\Phi+\Psi)}$. This can be obtained by correlating gravitational lensing with galaxy number counts, called galaxy-galaxy lensing~\cite{Prat:2017goa}. Observations of galaxy shapes provide a measurement of the convergence field $\kappa$
\be
\kappa(\bn,z)=\int_0^{r(z)}ds\frac{(r-s)s}{2r}\Delta_\perp(\Phi+\Psi)(\bn,s)\, . \label{kappa}
\ee
The Fourier transform of $\kappa(\bn,\tau)$ cannot be calculated in a straightforward way. It contains indeed an integral of $\kappa(\bn,\tau)$ on a hypersurface of constant time. However, the integral in Eq.~\eqref{kappa} is only meaningful on the past light-cone of the observer. Therefore, we first define the correlation function in configuration space, and then extract the power spectrum from this well-defined quantity. 

The observable $\Opsi$ depends on the bias difference between the bright and faint populations, $\bB-\bF$. In order to cancel this dependence we consider the following galaxy-galaxy lensing correlation
\begin{align}
&\xi^{\B\F}_{\Delta\kappa}\equiv\langle\Delta_\B(\bn,z)\kappa(\bn',z')\rangle-\langle\Delta_\F(\bn,z)\kappa(\bn',z')\rangle \\
&=(\bB-\bF)\int_0^{r'}ds \frac{(r'-s)s}{2r'}\langle\delta(\bn,z)\Delta_\perp (\Phi+\Psi)(\bn',s) \rangle\, ,\nonumber
\end{align}
where $r'\equiv r(z')$. Fourier transforming $\delta$ and $\Phi+\Psi$ and using Limber approximation~\cite{1953ApJ...117..134L,Kaiser:1996tp}, we obtain
\begin{align}
\xi^{\B\F}_{\Delta\kappa}&=-(\bB-\bF)\frac{(r'-r)r}{2r'}\Theta(r'-r)\\
&\times\frac{1}{2\pi}\int_0^\infty d k_\perp k^3_\perp P_{\delta(\Phi+\Psi)}(k_\perp,z)J_0(k_\perp\Delta x_\perp)\, .\nonumber
\end{align}
Here $J_0$ is the Bessel function of order zero, $k_\perp=|\bk_\perp|$ is the amplitude of the wave-number contribution perpendicular to the line-of-sight, $\Delta x_\perp=|\Delta \bx_\perp|$ denotes the amplitude of the vector joining the pixel in which $\Delta_\LL$ is measured and the pixel in which $\kappa$ is measured, projected in the plane orthogonal to the line-of-sight, and $\Theta$ is the Heaviside function, accounting for the fact that the correlation between $\Delta_\LL$ and $\kappa$ is non-zero only if $\kappa$ is behind $\Delta_\LL$. Note that $\langle\Delta_\LL\kappa\rangle$ contains also a lensing-lensing contribution, due to the second line in Eq.~\eqref{Delta}~\cite{Ghosh:2018nsm}. However this contribution does not depend on the galaxy population and vanishes therefore in $\xi^{\B\F}_{\Delta\kappa}$. Moreover, the correlation between $\kappa$ and the velocity contributions in Eq.~\eqref{Delta} exactly vanishes in the Limber approximation.

Since $\xi^{\B\F}_{\Delta\kappa}$ depends on $P_{\delta(\Phi+\Psi)}$, on the bias difference $\bB-\bF$ and on the observable quantities $r$ and $r'$, we could directly compare it with $\Opsi$ to extract the anisotropic stress. However, to build a more direct test, it is convenient to Fourier transform the correlation function and define
\begin{align}
&\Ophipsi(k_\perp,z)\equiv \label{Olens}\\
&\frac{-4\pi}{\Delta r k^2}\int_0^\infty d\Delta x_\perp \Delta x_\perp\xi^{\B\F}_{\Delta\kappa}(\Delta r, \Delta x_\perp,z)J_0(k_\perp\Delta x_\perp) \nonumber\\
&=(\bB-\bF)P_{\delta(\Phi+\Psi)}(k_\perp,z)\, .\nonumber
\end{align}
Here the correlation function is expressed in terms of the transverse separation, $\Delta x_\perp$, and the radial separation, $\Delta r$, between $\Delta_\LL$ and $\kappa$. To obtain the second equality in Eq.~\eqref{Olens} we have used the orthogonality relation for $J_0$.
Eq.~\eqref{Olens} contains an integral over $\Delta x_\perp$ going from 0 to $\infty$. In practice, since the correlation function and the Bessel function go to zero at large separation, the integral can be cut at some maximum transverse separation $\Delta x_\perp^{\rm max}$. The observable $\Ophipsi$ depends only on the transverse wave number $k_\perp$ (in the Limber approximation, the radial modes do not contribute to the correlation function $\xi^{\B\F}_{\Delta\kappa}$). Therefore, to compare with $\Opsi$, we have to evaluate $\Ophipsi$ at $k=k_\perp$.

The galaxy-galaxy lensing correlations have already been measured with high significance in several surveys, e.g.\ by the Dark Energy Survey~\cite{Prat:2017goa, DES:2017myr}. To build $\Ophipsi$ we need to measure these correlations for two different populations of galaxies and to take their difference. In a forthcoming paper we will study the SNR of this observable with the coming generation of surveys.

Note that $\Ophipsi$ has been build to cancel the bias difference of $\Opsi$, see Eq.~\eqref{Opsi}. This is however only possible if the galaxy-galaxy lensing correlations are measured from the same galaxy populations as the clustering correlations. It requires therefore a spectroscopic survey and a lensing survey that cover the same part of the sky~\cite{Cai_2012}. This is for example the case for Euclid, which will measure spectroscopic redshifts and photometric galaxy images over the same sky area~\cite{Euclid:2019clj}. Similarly, one can combine the spectroscopic redshifts measured by DESI~\cite{DESI:2016fyo}, with the galaxy images measured by the DESI Legacy Imaging Survey~\cite{Dey_2019}.

\section{Anisotropic stress estimator} 

\label{sec:stress}

In $\Lambda$CDM and for large classes of dark energy models, the two metric potentials are the same at late time, and we have therefore
\be
\Ostress=\frac{\Ophipsi}{\Opsi}=2\, .
\ee
On the other hand, in theories of modified gravity, the metric potentials are generally different. This difference, that can be parameterized by the variable $\eta$ through $\Phi(\bk,z)=\eta(k,z)\Psi(\bk,z)$, leads to 
\be
\Ostress=\frac{\Ophipsi}{\Opsi}=1+\eta\, .\label{anis}
\ee
The observable $\Opsi$ provides therefore a direct way of measuring $\eta$. In particular, if the ratio in Eq.~\eqref{anis} is different from 2 at any redshift or scale $k$, then $\Lambda$CDM is ruled out, as well as all classes of dark energy models with no anisotropic stress.

To emphasise the robustness of our test compared to standard methods, let us consider the following model, widely used in the literature: we parameterize deviations from GR by two functions, $\eta(k,z)$ (introduced above) and $Y(k,z)$, which encodes modifications to Poisson equation~\cite{Amendola:2012ky} 
\begin{align}
-k^2\Psi(\bk,z)=\frac{3}{2}\HH^2\Omega_m(z)Y(k,z)\delta(\bk,z)\, ,
\end{align}
where $\Omega_m(z)$ is the matter density parameter at redshift $z$. The function $Y(k,z)$ (sometimes called $\mu$ in the literature) reduces to 1 in GR. In addition to these functions, we allow for another departure from GR by modifying Euler equation
\be
\dot V(\bk,z)+\HH V(\bk, z)-k\Psi(\bk,z)=\Eb(\bk,z)\, ,
\ee
where $\Eb$ is a generic function encoding deviations from geodesic motion for dark matter. For example, in models where dark matter experiences a fifth force due to a non-minimal coupling to a scalar field, $\Eb$ takes the form
$\Eb=k\Gamma(z)\Psi(\bk,z)$, where $\Gamma$ is the amplitude of the fifth force~\cite{Bonvin:2018ckp}. 

We now apply the test developed in~\cite{Amendola:2012ky} to this particular model. By doing this we clearly use the test outside of its domain of validity, since in~\cite{Amendola:2012ky} it is explicitly assumed that Euler equation is valid. However, since in practice we do not know if dark matter obeys Euler equation or not, it is relevant to see what happens in this case. The evolution equation for the density contrast becomes
\begin{align}
\delta''+\left(2+\frac{H'}{H} \right)\delta'=-\frac{k^2}{(aH)^2}\Psi-\frac{k}{(aH)^2}\Eb\, ,    
\end{align}
where $H(z)=\HH(z)(1+z)$ and a prime denotes a derivative with respect to $N=\ln a$. The violation of the equivalence principle modifies therefore the way structures grow as a function of time, see also discussion in~\cite{Peebles:2002iq}. The combination of observables proposed in~\cite{Amendola:2012ky} to measure the anisotropic stress becomes then
\begin{align}
&\frac{3(1+z)^3P_2}{2E^2\left(P_3+2+\frac{E'}{E} \right)}-1=\label{testA}\\
&\eta+\frac{k}{(aH)^2}\frac{(1+\eta)}{f'+f^2+\left(2+\frac{E'}{E}\right)f}\frac{\Eb}{\delta}\neq \eta\, .\nonumber
\end{align}
Here $P_2$ and $P_3$ are the ratios of observables defined in~\cite{Amendola:2012ky} (see their Eqs. (14) and (15)), $f$ is the growth rate and $E\equiv H/H_0$. 

From Eq.~\eqref{testA} we see that the test developed in~\cite{Amendola:2012ky} is not a measurement of $\eta$ when Euler equation is not valid. In other words, a non-trivial outcome of this test can either mean that the anisotropic stress is non-zero, or that dark matter does not obey Euler equation.

\section{Conclusion}

\label{sec:conclusion}

In this paper, we have constructed an observable, $\Opsi$, which is directly proportional to the time component of the metric $\Psi$. This observable is constructed from the multipoles of the galaxy number counts, $\Delta$, and it relies only on observable quantities. We have then shown how this novel observable can be used to measure directly the anisotropic stress, i.e.\ the difference between the two metric potentials $\Phi$ and $\Psi$. 

This test, $\Ostress$, has the strong advantage that it does not rely on any assumption about the theory of gravity, apart from the fact that photons propagate on null geodesics. In particular, $\Ostress$ does not assume that dark matter obeys the continuity or Euler equation. This differs from standard measurement of the anisotropic stress, which rely on the validity of the continuity and Euler equations. As an example, we have shown how the test proposed in~\cite{Amendola:2012ky} will fail if Euler equation is not valid: instead of measuring directly $\eta$, the combination of observables defined in~\cite{Amendola:2012ky} contains an additional term proportional to the deviation from Euler equation. This limitation simply follows from the fact that standard observables are insensitive to $\Psi$. The only way to test the relation between $\Psi$ and $\Phi+\Psi$ is then to translate a measurement of $V$ into a measurement of $\Psi$ assuming that dark matter obeys Euler equation.

Our test, $\Ostress$, overcomes this limitation by using an observable sensitive to relativistic effects, which allows a direct measurement of $\Psi$. Of course, the price to pay is that $\Ostress$ will be more difficult to measure than standard tests, since relativistic effects are more challenging to measure than RSD. In a forthcoming paper we will study in more detail the sensitivity of $\Ostress$ for the coming generation of large-scale structure surveys, like DESI, Euclid and the SKA.  

Finally, let us note that our test, $\Ostress$, is highly complementary to the well-known $E_g$ statistics~\cite{Zhang:2007nk, Pullen:2015vtb, Ghosh:2018ijm}, which measures the ratio between density-lensing correlations, $\langle \delta(\Phi+\Psi)\rangle$, and density-velocity correlations, $\langle\delta V\rangle$. Similarly to our test, the $E_g$ statistics does not rely on the validity of Euler equation, and is therefore truly model-independent. It provides however constraints on the combination of parameters: $Y(1+\eta)/f$. An observed deviation from the $\Lambda$CDM value in $E_g$ can therefore either be due to a non-zero anisotropic stress (suggesting a modification of the theory of gravity) or to a growth rate which differs from $\Lambda$CDM (which also happens in simple models of dark energy). Having a test which directly and uniquely targets the anisotropic stress is therefore of high importance to test the theory of gravity.

\paragraph*{Acknowledgements.} This project has received funding from the European Research Council (ERC) under the European Union’s Horizon 2020 research and innovation programme (Grant agreement No. 863929; project title "Testing the law of gravity with novel large-scale structure observables"). We also acknowledge funding from the Swiss National Science Foundation. 


\bibliographystyle{apsrev4-1}
\bibliography{anisotropic.bib}

\begin{thebibliography}{59}%
\makeatletter
\providecommand \@ifxundefined [1]{%
 \@ifx{#1\undefined}
}%
\providecommand \@ifnum [1]{%
 \ifnum #1\expandafter \@firstoftwo
 \else \expandafter \@secondoftwo
 \fi
}%
\providecommand \@ifx [1]{%
 \ifx #1\expandafter \@firstoftwo
 \else \expandafter \@secondoftwo
 \fi
}%
\providecommand \natexlab [1]{#1}%
\providecommand \enquote  [1]{``#1''}%
\providecommand \bibnamefont  [1]{#1}%
\providecommand \bibfnamefont [1]{#1}%
\providecommand \citenamefont [1]{#1}%
\providecommand \href@noop [0]{\@secondoftwo}%
\providecommand \href [0]{\begingroup \@sanitize@url \@href}%
\providecommand \@href[1]{\@@startlink{#1}\@@href}%
\providecommand \@@href[1]{\endgroup#1\@@endlink}%
\providecommand \@sanitize@url [0]{\catcode `\\12\catcode `\$12\catcode
  `\&12\catcode `\#12\catcode `\^12\catcode `\_12\catcode `\%12\relax}%
\providecommand \@@startlink[1]{}%
\providecommand \@@endlink[0]{}%
\providecommand \url  [0]{\begingroup\@sanitize@url \@url }%
\providecommand \@url [1]{\endgroup\@href {#1}{\urlprefix }}%
\providecommand \urlprefix  [0]{URL }%
\providecommand \Eprint [0]{\href }%
\providecommand \doibase [0]{http://dx.doi.org/}%
\providecommand \selectlanguage [0]{\@gobble}%
\providecommand \bibinfo  [0]{\@secondoftwo}%
\providecommand \bibfield  [0]{\@secondoftwo}%
\providecommand \translation [1]{[#1]}%
\providecommand \BibitemOpen [0]{}%
\providecommand \bibitemStop [0]{}%
\providecommand \bibitemNoStop [0]{.\EOS\space}%
\providecommand \EOS [0]{\spacefactor3000\relax}%
\providecommand \BibitemShut  [1]{\csname bibitem#1\endcsname}%
\let\auto@bib@innerbib\@empty
\bibitem [{\citenamefont {Horndeski}(1974)}]{Horndeski1974}%
  \BibitemOpen
  \bibfield  {author} {\bibinfo {author} {\bibfnamefont {G.~W.}\ \bibnamefont
  {Horndeski}},\ }\href {\doibase 10.1007/BF01807638} {\bibfield  {journal}
  {\bibinfo  {journal} {International Journal of Theoretical Physics}\ }\textbf
  {\bibinfo {volume} {10}},\ \bibinfo {pages} {363} (\bibinfo {year}
  {1974})}\BibitemShut {NoStop}%
\bibitem [{\citenamefont {Zhang}\ \emph {et~al.}(2007)\citenamefont {Zhang},
  \citenamefont {Liguori}, \citenamefont {Bean},\ and\ \citenamefont
  {Dodelson}}]{Zhang:2007nk}%
  \BibitemOpen
  \bibfield  {author} {\bibinfo {author} {\bibfnamefont {P.}~\bibnamefont
  {Zhang}}, \bibinfo {author} {\bibfnamefont {M.}~\bibnamefont {Liguori}},
  \bibinfo {author} {\bibfnamefont {R.}~\bibnamefont {Bean}}, \ and\ \bibinfo
  {author} {\bibfnamefont {S.}~\bibnamefont {Dodelson}},\ }\href {\doibase
  10.1103/PhysRevLett.99.141302} {\bibfield  {journal} {\bibinfo  {journal}
  {Phys. Rev. Lett.}\ }\textbf {\bibinfo {volume} {99}},\ \bibinfo {pages}
  {141302} (\bibinfo {year} {2007})},\ \Eprint {http://arxiv.org/abs/0704.1932}
  {arXiv:0704.1932 [astro-ph]} \BibitemShut {NoStop}%
\bibitem [{\citenamefont {Amendola}\ \emph {et~al.}(2013)\citenamefont
  {Amendola}, \citenamefont {Kunz}, \citenamefont {Motta}, \citenamefont
  {Saltas},\ and\ \citenamefont {Sawicki}}]{Amendola:2012ky}%
  \BibitemOpen
  \bibfield  {author} {\bibinfo {author} {\bibfnamefont {L.}~\bibnamefont
  {Amendola}}, \bibinfo {author} {\bibfnamefont {M.}~\bibnamefont {Kunz}},
  \bibinfo {author} {\bibfnamefont {M.}~\bibnamefont {Motta}}, \bibinfo
  {author} {\bibfnamefont {I.~D.}\ \bibnamefont {Saltas}}, \ and\ \bibinfo
  {author} {\bibfnamefont {I.}~\bibnamefont {Sawicki}},\ }\href {\doibase
  10.1103/PhysRevD.87.023501} {\bibfield  {journal} {\bibinfo  {journal} {Phys.
  Rev. D}\ }\textbf {\bibinfo {volume} {87}},\ \bibinfo {pages} {023501}
  (\bibinfo {year} {2013})},\ \Eprint {http://arxiv.org/abs/1210.0439}
  {arXiv:1210.0439 [astro-ph.CO]} \BibitemShut {NoStop}%
\bibitem [{\citenamefont {Creminelli}\ \emph {et~al.}(2014)\citenamefont
  {Creminelli}, \citenamefont {Gleyzes}, \citenamefont {Hui}, \citenamefont
  {Simonovi\'c},\ and\ \citenamefont {Vernizzi}}]{Creminelli:2013nua}%
  \BibitemOpen
  \bibfield  {author} {\bibinfo {author} {\bibfnamefont {P.}~\bibnamefont
  {Creminelli}}, \bibinfo {author} {\bibfnamefont {J.}~\bibnamefont {Gleyzes}},
  \bibinfo {author} {\bibfnamefont {L.}~\bibnamefont {Hui}}, \bibinfo {author}
  {\bibfnamefont {M.}~\bibnamefont {Simonovi\'c}}, \ and\ \bibinfo {author}
  {\bibfnamefont {F.}~\bibnamefont {Vernizzi}},\ }\href {\doibase
  10.1088/1475-7516/2014/06/009} {\bibfield  {journal} {\bibinfo  {journal}
  {JCAP}\ }\textbf {\bibinfo {volume} {1406}},\ \bibinfo {pages} {009}
  (\bibinfo {year} {2014})},\ \Eprint {http://arxiv.org/abs/1312.6074}
  {arXiv:1312.6074 [astro-ph.CO]} \BibitemShut {NoStop}%
\bibitem [{\citenamefont {Kehagias}\ \emph {et~al.}(2014)\citenamefont
  {Kehagias}, \citenamefont {Noreña}, \citenamefont {Perrier},\ and\
  \citenamefont {Riotto}}]{Kehagias:2013rpa}%
  \BibitemOpen
  \bibfield  {author} {\bibinfo {author} {\bibfnamefont {A.}~\bibnamefont
  {Kehagias}}, \bibinfo {author} {\bibfnamefont {J.}~\bibnamefont {Noreña}},
  \bibinfo {author} {\bibfnamefont {H.}~\bibnamefont {Perrier}}, \ and\
  \bibinfo {author} {\bibfnamefont {A.}~\bibnamefont {Riotto}},\ }\href
  {\doibase 10.1016/j.nuclphysb.2014.03.020} {\bibfield  {journal} {\bibinfo
  {journal} {Nucl.\ Phys.\ B}\ }\textbf {\bibinfo {volume} {883}},\ \bibinfo
  {pages} {83} (\bibinfo {year} {2014})},\ \Eprint
  {http://arxiv.org/abs/1311.0786} {arXiv:1311.0786 [astro-ph.CO]} \BibitemShut
  {NoStop}%
\bibitem [{\citenamefont {Motta}\ \emph {et~al.}(2013)\citenamefont {Motta},
  \citenamefont {Sawicki}, \citenamefont {Saltas}, \citenamefont {Amendola},\
  and\ \citenamefont {Kunz}}]{Motta:2013cwa}%
  \BibitemOpen
  \bibfield  {author} {\bibinfo {author} {\bibfnamefont {M.}~\bibnamefont
  {Motta}}, \bibinfo {author} {\bibfnamefont {I.}~\bibnamefont {Sawicki}},
  \bibinfo {author} {\bibfnamefont {I.~D.}\ \bibnamefont {Saltas}}, \bibinfo
  {author} {\bibfnamefont {L.}~\bibnamefont {Amendola}}, \ and\ \bibinfo
  {author} {\bibfnamefont {M.}~\bibnamefont {Kunz}},\ }\href {\doibase
  10.1103/PhysRevD.88.124035} {\bibfield  {journal} {\bibinfo  {journal} {Phys.
  Rev. D}\ }\textbf {\bibinfo {volume} {88}},\ \bibinfo {pages} {124035}
  (\bibinfo {year} {2013})},\ \Eprint {http://arxiv.org/abs/1305.0008}
  {arXiv:1305.0008 [astro-ph.CO]} \BibitemShut {NoStop}%
\bibitem [{\citenamefont {Ghosh}\ and\ \citenamefont
  {Durrer}(2019)}]{Ghosh:2018ijm}%
  \BibitemOpen
  \bibfield  {author} {\bibinfo {author} {\bibfnamefont {B.}~\bibnamefont
  {Ghosh}}\ and\ \bibinfo {author} {\bibfnamefont {R.}~\bibnamefont {Durrer}},\
  }\href {\doibase 10.1088/1475-7516/2019/06/010} {\bibfield  {journal}
  {\bibinfo  {journal} {JCAP}\ }\textbf {\bibinfo {volume} {06}},\ \bibinfo
  {pages} {010} (\bibinfo {year} {2019})},\ \Eprint
  {http://arxiv.org/abs/1812.09546} {arXiv:1812.09546 [astro-ph.CO]}
  \BibitemShut {NoStop}%
\bibitem [{\citenamefont {Bonvin}\ and\ \citenamefont
  {Fleury}(2018)}]{Bonvin:2018ckp}%
  \BibitemOpen
  \bibfield  {author} {\bibinfo {author} {\bibfnamefont {C.}~\bibnamefont
  {Bonvin}}\ and\ \bibinfo {author} {\bibfnamefont {P.}~\bibnamefont
  {Fleury}},\ }\href {\doibase 10.1088/1475-7516/2018/05/061} {\bibfield
  {journal} {\bibinfo  {journal} {JCAP}\ }\textbf {\bibinfo {volume} {1805}},\
  \bibinfo {pages} {061} (\bibinfo {year} {2018})},\ \Eprint
  {http://arxiv.org/abs/1803.02771} {arXiv:1803.02771 [astro-ph.CO]}
  \BibitemShut {NoStop}%
\bibitem [{\citenamefont {Franco}\ \emph {et~al.}(2019)\citenamefont {Franco},
  \citenamefont {Bonvin},\ and\ \citenamefont {Clarkson}}]{Franco:2019wbj}%
  \BibitemOpen
  \bibfield  {author} {\bibinfo {author} {\bibfnamefont {F.~O.}\ \bibnamefont
  {Franco}}, \bibinfo {author} {\bibfnamefont {C.}~\bibnamefont {Bonvin}}, \
  and\ \bibinfo {author} {\bibfnamefont {C.}~\bibnamefont {Clarkson}},\
  }\href@noop {} {\  (\bibinfo {year} {2019})},\ \Eprint
  {http://arxiv.org/abs/1906.02217} {arXiv:1906.02217 [astro-ph.CO]}
  \BibitemShut {NoStop}%
\bibitem [{\citenamefont {Bonvin}\ \emph {et~al.}(2020)\citenamefont {Bonvin},
  \citenamefont {Franco},\ and\ \citenamefont {Fleury}}]{Bonvin:2020cxp}%
  \BibitemOpen
  \bibfield  {author} {\bibinfo {author} {\bibfnamefont {C.}~\bibnamefont
  {Bonvin}}, \bibinfo {author} {\bibfnamefont {F.~O.}\ \bibnamefont {Franco}},
  \ and\ \bibinfo {author} {\bibfnamefont {P.}~\bibnamefont {Fleury}},\ }\href
  {\doibase 10.1088/1475-7516/2020/08/004} {\bibfield  {journal} {\bibinfo
  {journal} {JCAP}\ }\textbf {\bibinfo {volume} {08}},\ \bibinfo {pages} {004}
  (\bibinfo {year} {2020})},\ \Eprint {http://arxiv.org/abs/2004.06457}
  {arXiv:2004.06457 [astro-ph.CO]} \BibitemShut {NoStop}%
\bibitem [{\citenamefont {Parkinson}\ \emph {et~al.}(2012)\citenamefont
  {Parkinson} \emph {et~al.}}]{Parkinson:2012vd}%
  \BibitemOpen
  \bibfield  {author} {\bibinfo {author} {\bibfnamefont {D.}~\bibnamefont
  {Parkinson}} \emph {et~al.},\ }\href {\doibase 10.1103/PhysRevD.86.103518}
  {\bibfield  {journal} {\bibinfo  {journal} {Phys. Rev. D}\ }\textbf {\bibinfo
  {volume} {86}},\ \bibinfo {pages} {103518} (\bibinfo {year} {2012})},\
  \Eprint {http://arxiv.org/abs/1210.2130} {arXiv:1210.2130 [astro-ph.CO]}
  \BibitemShut {NoStop}%
\bibitem [{\citenamefont {Alam}\ \emph {et~al.}(2017)\citenamefont {Alam} \emph
  {et~al.}}]{Alam:2016hwk}%
  \BibitemOpen
  \bibfield  {author} {\bibinfo {author} {\bibfnamefont {S.}~\bibnamefont
  {Alam}} \emph {et~al.} (\bibinfo {collaboration} {BOSS}),\ }\href {\doibase
  10.1093/mnras/stx721} {\bibfield  {journal} {\bibinfo  {journal} {Mon. Not.
  Roy. Astron. Soc.}\ }\textbf {\bibinfo {volume} {470}},\ \bibinfo {pages}
  {2617} (\bibinfo {year} {2017})},\ \Eprint {http://arxiv.org/abs/1607.03155}
  {arXiv:1607.03155 [astro-ph.CO]} \BibitemShut {NoStop}%
\bibitem [{\citenamefont {Asgari}\ \emph {et~al.}(2021)\citenamefont {Asgari}
  \emph {et~al.}}]{Asgari:2020wuj}%
  \BibitemOpen
  \bibfield  {author} {\bibinfo {author} {\bibfnamefont {M.}~\bibnamefont
  {Asgari}} \emph {et~al.} (\bibinfo {collaboration} {KiDS}),\ }\href {\doibase
  10.1051/0004-6361/202039070} {\bibfield  {journal} {\bibinfo  {journal}
  {Astron. Astrophys.}\ }\textbf {\bibinfo {volume} {645}},\ \bibinfo {pages}
  {A104} (\bibinfo {year} {2021})},\ \Eprint {http://arxiv.org/abs/2007.15633}
  {arXiv:2007.15633 [astro-ph.CO]} \BibitemShut {NoStop}%
\bibitem [{\citenamefont {Gatti}\ \emph {et~al.}(2020)\citenamefont {Gatti}
  \emph {et~al.}}]{Gatti:2020eyt}%
  \BibitemOpen
  \bibfield  {author} {\bibinfo {author} {\bibfnamefont {M.}~\bibnamefont
  {Gatti}} \emph {et~al.} (\bibinfo {collaboration} {DES}),\ }\href@noop {} {\
  (\bibinfo {year} {2020})},\ \Eprint {http://arxiv.org/abs/2011.03408}
  {arXiv:2011.03408 [astro-ph.CO]} \BibitemShut {NoStop}%
\bibitem [{\citenamefont {Sherwin}\ \emph {et~al.}(2017)\citenamefont {Sherwin}
  \emph {et~al.}}]{Sherwin:2016tyf}%
  \BibitemOpen
  \bibfield  {author} {\bibinfo {author} {\bibfnamefont {B.~D.}\ \bibnamefont
  {Sherwin}} \emph {et~al.},\ }\href {\doibase 10.1103/PhysRevD.95.123529}
  {\bibfield  {journal} {\bibinfo  {journal} {Phys. Rev. D}\ }\textbf {\bibinfo
  {volume} {95}},\ \bibinfo {pages} {123529} (\bibinfo {year} {2017})},\
  \Eprint {http://arxiv.org/abs/1611.09753} {arXiv:1611.09753 [astro-ph.CO]}
  \BibitemShut {NoStop}%
\bibitem [{\citenamefont {Aghanim}\ \emph {et~al.}(2020)\citenamefont {Aghanim}
  \emph {et~al.}}]{Aghanim:2018oex}%
  \BibitemOpen
  \bibfield  {author} {\bibinfo {author} {\bibfnamefont {N.}~\bibnamefont
  {Aghanim}} \emph {et~al.} (\bibinfo {collaboration} {Planck}),\ }\href
  {\doibase 10.1051/0004-6361/201833886} {\bibfield  {journal} {\bibinfo
  {journal} {Astron. Astrophys.}\ }\textbf {\bibinfo {volume} {641}},\ \bibinfo
  {pages} {A8} (\bibinfo {year} {2020})},\ \Eprint
  {http://arxiv.org/abs/1807.06210} {arXiv:1807.06210 [astro-ph.CO]}
  \BibitemShut {NoStop}%
\bibitem [{\citenamefont {Wu}\ \emph {et~al.}(2019)\citenamefont {Wu} \emph
  {et~al.}}]{Wu:2019hek}%
  \BibitemOpen
  \bibfield  {author} {\bibinfo {author} {\bibfnamefont {W.~L.~K.}\
  \bibnamefont {Wu}} \emph {et~al.},\ }\href {\doibase
  10.3847/1538-4357/ab4186} {\bibfield  {journal} {\bibinfo  {journal}
  {Astrophys. J.}\ }\textbf {\bibinfo {volume} {884}},\ \bibinfo {pages} {70}
  (\bibinfo {year} {2019})},\ \Eprint {http://arxiv.org/abs/1905.05777}
  {arXiv:1905.05777 [astro-ph.CO]} \BibitemShut {NoStop}%
\bibitem [{\citenamefont {Aguilar~Fa\'undez}\ \emph {et~al.}(2020)\citenamefont
  {Aguilar~Fa\'undez} \emph {et~al.}}]{Faundez:2019lmz}%
  \BibitemOpen
  \bibfield  {author} {\bibinfo {author} {\bibfnamefont {M.~A.~O.}\
  \bibnamefont {Aguilar~Fa\'undez}} \emph {et~al.} (\bibinfo {collaboration}
  {POLARBEAR}),\ }\href {\doibase 10.3847/1538-4357/ab7e29} {\bibfield
  {journal} {\bibinfo  {journal} {Astrophys. J.}\ }\textbf {\bibinfo {volume}
  {893}},\ \bibinfo {pages} {85} (\bibinfo {year} {2020})},\ \Eprint
  {http://arxiv.org/abs/1911.10980} {arXiv:1911.10980 [astro-ph.CO]}
  \BibitemShut {NoStop}%
\bibitem [{\citenamefont {Ade}\ \emph {et~al.}(2016)\citenamefont {Ade} \emph
  {et~al.}}]{Ade:2015dva}%
  \BibitemOpen
  \bibfield  {author} {\bibinfo {author} {\bibfnamefont {P.~A.~R.}\
  \bibnamefont {Ade}} \emph {et~al.} (\bibinfo {collaboration} {Planck}),\
  }\href {\doibase 10.1051/0004-6361/201525831} {\bibfield  {journal} {\bibinfo
   {journal} {Astron. Astrophys.}\ }\textbf {\bibinfo {volume} {594}},\
  \bibinfo {pages} {A21} (\bibinfo {year} {2016})},\ \Eprint
  {http://arxiv.org/abs/1502.01595} {arXiv:1502.01595 [astro-ph.CO]}
  \BibitemShut {NoStop}%
\bibitem [{\citenamefont {Amendola}\ \emph {et~al.}(2014)\citenamefont
  {Amendola}, \citenamefont {Fogli}, \citenamefont {Guarnizo}, \citenamefont
  {Kunz},\ and\ \citenamefont {Vollmer}}]{Amendola:2013qna}%
  \BibitemOpen
  \bibfield  {author} {\bibinfo {author} {\bibfnamefont {L.}~\bibnamefont
  {Amendola}}, \bibinfo {author} {\bibfnamefont {S.}~\bibnamefont {Fogli}},
  \bibinfo {author} {\bibfnamefont {A.}~\bibnamefont {Guarnizo}}, \bibinfo
  {author} {\bibfnamefont {M.}~\bibnamefont {Kunz}}, \ and\ \bibinfo {author}
  {\bibfnamefont {A.}~\bibnamefont {Vollmer}},\ }\href {\doibase
  10.1103/PhysRevD.89.063538} {\bibfield  {journal} {\bibinfo  {journal} {Phys.
  Rev. D}\ }\textbf {\bibinfo {volume} {89}},\ \bibinfo {pages} {063538}
  (\bibinfo {year} {2014})},\ \Eprint {http://arxiv.org/abs/1311.4765}
  {arXiv:1311.4765 [astro-ph.CO]} \BibitemShut {NoStop}%
\bibitem [{\citenamefont {Pinho}\ \emph {et~al.}(2018)\citenamefont {Pinho},
  \citenamefont {Casas},\ and\ \citenamefont {Amendola}}]{Pinho:2018unz}%
  \BibitemOpen
  \bibfield  {author} {\bibinfo {author} {\bibfnamefont {A.~M.}\ \bibnamefont
  {Pinho}}, \bibinfo {author} {\bibfnamefont {S.}~\bibnamefont {Casas}}, \ and\
  \bibinfo {author} {\bibfnamefont {L.}~\bibnamefont {Amendola}},\ }\href
  {\doibase 10.1088/1475-7516/2018/11/027} {\bibfield  {journal} {\bibinfo
  {journal} {JCAP}\ }\textbf {\bibinfo {volume} {11}},\ \bibinfo {pages} {027}
  (\bibinfo {year} {2018})},\ \Eprint {http://arxiv.org/abs/1805.00027}
  {arXiv:1805.00027 [astro-ph.CO]} \BibitemShut {NoStop}%
\bibitem [{\citenamefont {Arjona}\ and\ \citenamefont
  {Nesseris}(2020)}]{Arjona:2020kco}%
  \BibitemOpen
  \bibfield  {author} {\bibinfo {author} {\bibfnamefont {R.}~\bibnamefont
  {Arjona}}\ and\ \bibinfo {author} {\bibfnamefont {S.}~\bibnamefont
  {Nesseris}},\ }\href {\doibase 10.1088/1475-7516/2020/11/042} {\bibfield
  {journal} {\bibinfo  {journal} {JCAP}\ }\textbf {\bibinfo {volume} {11}},\
  \bibinfo {pages} {042} (\bibinfo {year} {2020})},\ \Eprint
  {http://arxiv.org/abs/2001.11420} {arXiv:2001.11420 [astro-ph.CO]}
  \BibitemShut {NoStop}%
\bibitem [{\citenamefont {Adamek}\ \emph {et~al.}(2016)\citenamefont {Adamek},
  \citenamefont {Daverio}, \citenamefont {Durrer},\ and\ \citenamefont
  {Kunz}}]{Adamek:2016zes}%
  \BibitemOpen
  \bibfield  {author} {\bibinfo {author} {\bibfnamefont {J.}~\bibnamefont
  {Adamek}}, \bibinfo {author} {\bibfnamefont {D.}~\bibnamefont {Daverio}},
  \bibinfo {author} {\bibfnamefont {R.}~\bibnamefont {Durrer}}, \ and\ \bibinfo
  {author} {\bibfnamefont {M.}~\bibnamefont {Kunz}},\ }\href {\doibase
  10.1088/1475-7516/2016/07/053} {\bibfield  {journal} {\bibinfo  {journal}
  {JCAP}\ }\textbf {\bibinfo {volume} {07}},\ \bibinfo {pages} {053} (\bibinfo
  {year} {2016})},\ \Eprint {http://arxiv.org/abs/1604.06065} {arXiv:1604.06065
  [astro-ph.CO]} \BibitemShut {NoStop}%
\bibitem [{\citenamefont {Adamek}\ \emph {et~al.}(2017)\citenamefont {Adamek},
  \citenamefont {Durrer},\ and\ \citenamefont {Kunz}}]{Adamek:2017uiq}%
  \BibitemOpen
  \bibfield  {author} {\bibinfo {author} {\bibfnamefont {J.}~\bibnamefont
  {Adamek}}, \bibinfo {author} {\bibfnamefont {R.}~\bibnamefont {Durrer}}, \
  and\ \bibinfo {author} {\bibfnamefont {M.}~\bibnamefont {Kunz}},\ }\href
  {\doibase 10.1088/1475-7516/2017/11/004} {\bibfield  {journal} {\bibinfo
  {journal} {JCAP}\ }\textbf {\bibinfo {volume} {11}},\ \bibinfo {pages} {004}
  (\bibinfo {year} {2017})},\ \Eprint {http://arxiv.org/abs/1707.06938}
  {arXiv:1707.06938 [astro-ph.CO]} \BibitemShut {NoStop}%
\bibitem [{\citenamefont {Saltas}\ \emph {et~al.}(2014)\citenamefont {Saltas},
  \citenamefont {Sawicki}, \citenamefont {Amendola},\ and\ \citenamefont
  {Kunz}}]{Saltas:2014dha}%
  \BibitemOpen
  \bibfield  {author} {\bibinfo {author} {\bibfnamefont {I.~D.}\ \bibnamefont
  {Saltas}}, \bibinfo {author} {\bibfnamefont {I.}~\bibnamefont {Sawicki}},
  \bibinfo {author} {\bibfnamefont {L.}~\bibnamefont {Amendola}}, \ and\
  \bibinfo {author} {\bibfnamefont {M.}~\bibnamefont {Kunz}},\ }\href {\doibase
  10.1103/PhysRevLett.113.191101} {\bibfield  {journal} {\bibinfo  {journal}
  {Phys. Rev. Lett.}\ }\textbf {\bibinfo {volume} {113}},\ \bibinfo {pages}
  {191101} (\bibinfo {year} {2014})},\ \Eprint {http://arxiv.org/abs/1406.7139}
  {arXiv:1406.7139 [astro-ph.CO]} \BibitemShut {NoStop}%
\bibitem [{\citenamefont {Yoo}\ \emph {et~al.}(2009)\citenamefont {Yoo},
  \citenamefont {Fitzpatrick},\ and\ \citenamefont {Zaldarriaga}}]{Yoo:2009au}%
  \BibitemOpen
  \bibfield  {author} {\bibinfo {author} {\bibfnamefont {J.}~\bibnamefont
  {Yoo}}, \bibinfo {author} {\bibfnamefont {A.~L.}\ \bibnamefont
  {Fitzpatrick}}, \ and\ \bibinfo {author} {\bibfnamefont {M.}~\bibnamefont
  {Zaldarriaga}},\ }\href {\doibase 10.1103/PhysRevD.80.083514} {\bibfield
  {journal} {\bibinfo  {journal} {Phys. Rev.}\ }\textbf {\bibinfo {volume}
  {D80}},\ \bibinfo {pages} {083514} (\bibinfo {year} {2009})},\ \Eprint
  {http://arxiv.org/abs/0907.0707} {arXiv:0907.0707 [astro-ph.CO]} \BibitemShut
  {NoStop}%
\bibitem [{\citenamefont {Bonvin}\ and\ \citenamefont
  {Durrer}(2011)}]{Bonvin:2011bg}%
  \BibitemOpen
  \bibfield  {author} {\bibinfo {author} {\bibfnamefont {C.}~\bibnamefont
  {Bonvin}}\ and\ \bibinfo {author} {\bibfnamefont {R.}~\bibnamefont
  {Durrer}},\ }\href {\doibase 10.1103/PhysRevD.84.063505} {\bibfield
  {journal} {\bibinfo  {journal} {Phys. Rev.}\ }\textbf {\bibinfo {volume}
  {D84}},\ \bibinfo {pages} {063505} (\bibinfo {year} {2011})},\ \Eprint
  {http://arxiv.org/abs/1105.5280} {arXiv:1105.5280 [astro-ph.CO]} \BibitemShut
  {NoStop}%
\bibitem [{\citenamefont {Challinor}\ and\ \citenamefont
  {Lewis}(2011)}]{Challinor:2011bk}%
  \BibitemOpen
  \bibfield  {author} {\bibinfo {author} {\bibfnamefont {A.}~\bibnamefont
  {Challinor}}\ and\ \bibinfo {author} {\bibfnamefont {A.}~\bibnamefont
  {Lewis}},\ }\href {\doibase 10.1103/PhysRevD.84.043516} {\bibfield  {journal}
  {\bibinfo  {journal} {Phys. Rev.}\ }\textbf {\bibinfo {volume} {D84}},\
  \bibinfo {pages} {043516} (\bibinfo {year} {2011})},\ \Eprint
  {http://arxiv.org/abs/1105.5292} {arXiv:1105.5292 [astro-ph.CO]} \BibitemShut
  {NoStop}%
\bibitem [{\citenamefont {Jeong}\ \emph {et~al.}(2012)\citenamefont {Jeong},
  \citenamefont {Schmidt},\ and\ \citenamefont {Hirata}}]{Jeong:2011as}%
  \BibitemOpen
  \bibfield  {author} {\bibinfo {author} {\bibfnamefont {D.}~\bibnamefont
  {Jeong}}, \bibinfo {author} {\bibfnamefont {F.}~\bibnamefont {Schmidt}}, \
  and\ \bibinfo {author} {\bibfnamefont {C.~M.}\ \bibnamefont {Hirata}},\
  }\href {\doibase 10.1103/PhysRevD.85.023504} {\bibfield  {journal} {\bibinfo
  {journal} {Phys. Rev. D}\ }\textbf {\bibinfo {volume} {85}},\ \bibinfo
  {pages} {023504} (\bibinfo {year} {2012})},\ \Eprint
  {http://arxiv.org/abs/1107.5427} {arXiv:1107.5427 [astro-ph.CO]} \BibitemShut
  {NoStop}%
\bibitem [{\citenamefont {Kaiser}(1987)}]{Kaiser:1987qv}%
  \BibitemOpen
  \bibfield  {author} {\bibinfo {author} {\bibfnamefont {N.}~\bibnamefont
  {Kaiser}},\ }\href@noop {} {\bibfield  {journal} {\bibinfo  {journal} {Mon.
  Not. Roy. Astron. Soc.}\ }\textbf {\bibinfo {volume} {227}},\ \bibinfo
  {pages} {1} (\bibinfo {year} {1987})}\BibitemShut {NoStop}%
\bibitem [{\citenamefont {Hamilton}(1997)}]{Hamilton:1997zq}%
  \BibitemOpen
  \bibfield  {author} {\bibinfo {author} {\bibfnamefont {A.~J.~S.}\
  \bibnamefont {Hamilton}},\ }in\ \href {\doibase 10.1007/978-94-011-4960-0_17}
  {\emph {\bibinfo {booktitle} {{Ringberg Workshop on Large Scale Structure
  Ringberg, Germany, September 23-28, 1996}}}}\ (\bibinfo {year} {1997})\
  \Eprint {http://arxiv.org/abs/astro-ph/9708102} {arXiv:astro-ph/9708102
  [astro-ph]} \BibitemShut {NoStop}%
\bibitem [{\citenamefont {Scranton}\ \emph {et~al.}(2005)\citenamefont
  {Scranton} \emph {et~al.}}]{Scranton:2005ci}%
  \BibitemOpen
  \bibfield  {author} {\bibinfo {author} {\bibfnamefont {R.}~\bibnamefont
  {Scranton}} \emph {et~al.} (\bibinfo {collaboration} {SDSS}),\ }\href
  {\doibase 10.1086/431358} {\bibfield  {journal} {\bibinfo  {journal}
  {Astrophys. J.}\ }\textbf {\bibinfo {volume} {633}},\ \bibinfo {pages} {589}
  (\bibinfo {year} {2005})},\ \Eprint {http://arxiv.org/abs/astro-ph/0504510}
  {arXiv:astro-ph/0504510} \BibitemShut {NoStop}%
\bibitem [{\citenamefont {Garcia-Fernandez}\ \emph {et~al.}(2018)\citenamefont
  {Garcia-Fernandez} \emph {et~al.}}]{Garcia-Fernandez:2016oud}%
  \BibitemOpen
  \bibfield  {author} {\bibinfo {author} {\bibfnamefont {M.}~\bibnamefont
  {Garcia-Fernandez}} \emph {et~al.} (\bibinfo {collaboration} {DES}),\ }\href
  {\doibase 10.1093/mnras/sty282} {\bibfield  {journal} {\bibinfo  {journal}
  {Mon. Not. Roy. Astron. Soc.}\ }\textbf {\bibinfo {volume} {476}},\ \bibinfo
  {pages} {1071} (\bibinfo {year} {2018})},\ \Eprint
  {http://arxiv.org/abs/1611.10326} {arXiv:1611.10326 [astro-ph.CO]}
  \BibitemShut {NoStop}%
\bibitem [{\citenamefont {{Bonvin}}\ \emph {et~al.}(2014)\citenamefont
  {{Bonvin}}, \citenamefont {{Hui}},\ and\ \citenamefont
  {{Gazta{\~n}aga}}}]{2014PhRvD..89h3535B}%
  \BibitemOpen
  \bibfield  {author} {\bibinfo {author} {\bibfnamefont {C.}~\bibnamefont
  {{Bonvin}}}, \bibinfo {author} {\bibfnamefont {L.}~\bibnamefont {{Hui}}}, \
  and\ \bibinfo {author} {\bibfnamefont {E.}~\bibnamefont {{Gazta{\~n}aga}}},\
  }\href {\doibase 10.1103/PhysRevD.89.083535} {\bibfield  {journal} {\bibinfo
  {journal} {\prd}\ }\textbf {\bibinfo {volume} {89}},\ \bibinfo {eid} {083535}
  (\bibinfo {year} {2014})},\ \Eprint {http://arxiv.org/abs/1309.1321}
  {arXiv:1309.1321} \BibitemShut {NoStop}%
\bibitem [{\citenamefont {Jelic-Cizmek}\ \emph {et~al.}(2020)\citenamefont
  {Jelic-Cizmek}, \citenamefont {Lepori}, \citenamefont {Bonvin},\ and\
  \citenamefont {Durrer}}]{Jelic-Cizmek:2020pkh}%
  \BibitemOpen
  \bibfield  {author} {\bibinfo {author} {\bibfnamefont {G.}~\bibnamefont
  {Jelic-Cizmek}}, \bibinfo {author} {\bibfnamefont {F.}~\bibnamefont
  {Lepori}}, \bibinfo {author} {\bibfnamefont {C.}~\bibnamefont {Bonvin}}, \
  and\ \bibinfo {author} {\bibfnamefont {R.}~\bibnamefont {Durrer}},\
  }\href@noop {} {\  (\bibinfo {year} {2020})},\ \Eprint
  {http://arxiv.org/abs/2004.12981} {arXiv:2004.12981 [astro-ph.CO]}
  \BibitemShut {NoStop}%
\bibitem [{\citenamefont {McDonald}(2009)}]{McDonald:2009ud}%
  \BibitemOpen
  \bibfield  {author} {\bibinfo {author} {\bibfnamefont {P.}~\bibnamefont
  {McDonald}},\ }\href {\doibase 10.1088/1475-7516/2009/11/026} {\bibfield
  {journal} {\bibinfo  {journal} {JCAP}\ }\textbf {\bibinfo {volume} {0911}},\
  \bibinfo {pages} {026} (\bibinfo {year} {2009})},\ \Eprint
  {http://arxiv.org/abs/0907.5220} {arXiv:0907.5220 [astro-ph.CO]} \BibitemShut
  {NoStop}%
\bibitem [{\citenamefont {Yoo}\ \emph {et~al.}(2012)\citenamefont {Yoo},
  \citenamefont {Hamaus}, \citenamefont {Seljak},\ and\ \citenamefont
  {Zaldarriaga}}]{Yoo:2012se}%
  \BibitemOpen
  \bibfield  {author} {\bibinfo {author} {\bibfnamefont {J.}~\bibnamefont
  {Yoo}}, \bibinfo {author} {\bibfnamefont {N.}~\bibnamefont {Hamaus}},
  \bibinfo {author} {\bibfnamefont {U.}~\bibnamefont {Seljak}}, \ and\ \bibinfo
  {author} {\bibfnamefont {M.}~\bibnamefont {Zaldarriaga}},\ }\href {\doibase
  10.1103/PhysRevD.86.063514} {\bibfield  {journal} {\bibinfo  {journal} {Phys.
  Rev.}\ }\textbf {\bibinfo {volume} {D86}},\ \bibinfo {pages} {063514}
  (\bibinfo {year} {2012})},\ \Eprint {http://arxiv.org/abs/1206.5809}
  {arXiv:1206.5809 [astro-ph.CO]} \BibitemShut {NoStop}%
\bibitem [{\citenamefont {Croft}(2013)}]{Croft:2013taa}%
  \BibitemOpen
  \bibfield  {author} {\bibinfo {author} {\bibfnamefont {R.~A.~C.}\
  \bibnamefont {Croft}},\ }\href {\doibase 10.1093/mnras/stt1223} {\bibfield
  {journal} {\bibinfo  {journal} {Mon. Not. Roy. Astron. Soc.}\ }\textbf
  {\bibinfo {volume} {434}},\ \bibinfo {pages} {3008} (\bibinfo {year}
  {2013})},\ \Eprint {http://arxiv.org/abs/1304.4124} {arXiv:1304.4124
  [astro-ph.CO]} \BibitemShut {NoStop}%
\bibitem [{\citenamefont {Scoccimarro}\ \emph {et~al.}(1999)\citenamefont
  {Scoccimarro}, \citenamefont {Couchman},\ and\ \citenamefont
  {Frieman}}]{Scoccimarro:1999ed}%
  \BibitemOpen
  \bibfield  {author} {\bibinfo {author} {\bibfnamefont {R.}~\bibnamefont
  {Scoccimarro}}, \bibinfo {author} {\bibfnamefont {H.~M.~P.}\ \bibnamefont
  {Couchman}}, \ and\ \bibinfo {author} {\bibfnamefont {J.~A.}\ \bibnamefont
  {Frieman}},\ }\href {\doibase 10.1086/307220} {\bibfield  {journal} {\bibinfo
   {journal} {Astrophys. J.}\ }\textbf {\bibinfo {volume} {517}},\ \bibinfo
  {pages} {531} (\bibinfo {year} {1999})},\ \Eprint
  {http://arxiv.org/abs/astro-ph/9808305} {arXiv:astro-ph/9808305} \BibitemShut
  {NoStop}%
\bibitem [{\citenamefont {Nielsen}\ and\ \citenamefont
  {Durrer}(2017)}]{Nielsen:2016ldx}%
  \BibitemOpen
  \bibfield  {author} {\bibinfo {author} {\bibfnamefont {J.~T.}\ \bibnamefont
  {Nielsen}}\ and\ \bibinfo {author} {\bibfnamefont {R.}~\bibnamefont
  {Durrer}},\ }\href {\doibase 10.1088/1475-7516/2017/03/010} {\bibfield
  {journal} {\bibinfo  {journal} {JCAP}\ }\textbf {\bibinfo {volume} {03}},\
  \bibinfo {pages} {010} (\bibinfo {year} {2017})},\ \Eprint
  {http://arxiv.org/abs/1606.02113} {arXiv:1606.02113 [astro-ph.CO]}
  \BibitemShut {NoStop}%
\bibitem [{\citenamefont {Zhao}\ \emph {et~al.}(2013)\citenamefont {Zhao},
  \citenamefont {Peacock},\ and\ \citenamefont {Li}}]{Zhao:2012gxk}%
  \BibitemOpen
  \bibfield  {author} {\bibinfo {author} {\bibfnamefont {H.}~\bibnamefont
  {Zhao}}, \bibinfo {author} {\bibfnamefont {J.~A.}\ \bibnamefont {Peacock}}, \
  and\ \bibinfo {author} {\bibfnamefont {B.}~\bibnamefont {Li}},\ }\href
  {\doibase 10.1103/PhysRevD.88.043013} {\bibfield  {journal} {\bibinfo
  {journal} {Phys. Rev. D}\ }\textbf {\bibinfo {volume} {88}},\ \bibinfo
  {pages} {043013} (\bibinfo {year} {2013})},\ \Eprint
  {http://arxiv.org/abs/1206.5032} {arXiv:1206.5032 [astro-ph.CO]} \BibitemShut
  {NoStop}%
\bibitem [{\citenamefont {Kaiser}(2013)}]{Kaiser:2013ipa}%
  \BibitemOpen
  \bibfield  {author} {\bibinfo {author} {\bibfnamefont {N.}~\bibnamefont
  {Kaiser}},\ }\href {\doibase 10.1093/mnras/stt1370} {\bibfield  {journal}
  {\bibinfo  {journal} {Mon. Not. Roy. Astron. Soc.}\ }\textbf {\bibinfo
  {volume} {435}},\ \bibinfo {pages} {1278} (\bibinfo {year} {2013})},\ \Eprint
  {http://arxiv.org/abs/1303.3663} {arXiv:1303.3663 [astro-ph.CO]} \BibitemShut
  {NoStop}%
\bibitem [{\citenamefont {Cai}\ \emph {et~al.}(2017)\citenamefont {Cai},
  \citenamefont {Kaiser}, \citenamefont {Cole},\ and\ \citenamefont
  {Frenk}}]{Cai:2016ors}%
  \BibitemOpen
  \bibfield  {author} {\bibinfo {author} {\bibfnamefont {Y.-C.}\ \bibnamefont
  {Cai}}, \bibinfo {author} {\bibfnamefont {N.}~\bibnamefont {Kaiser}},
  \bibinfo {author} {\bibfnamefont {S.}~\bibnamefont {Cole}}, \ and\ \bibinfo
  {author} {\bibfnamefont {C.}~\bibnamefont {Frenk}},\ }\href {\doibase
  10.1093/mnras/stx469} {\bibfield  {journal} {\bibinfo  {journal} {Mon. Not.
  Roy. Astron. Soc.}\ }\textbf {\bibinfo {volume} {468}},\ \bibinfo {pages}
  {1981} (\bibinfo {year} {2017})},\ \Eprint {http://arxiv.org/abs/1609.04864}
  {arXiv:1609.04864 [astro-ph.CO]} \BibitemShut {NoStop}%
\bibitem [{\citenamefont {Beutler}\ and\ \citenamefont
  {Di~Dio}(2020)}]{Beutler:2020evf}%
  \BibitemOpen
  \bibfield  {author} {\bibinfo {author} {\bibfnamefont {F.}~\bibnamefont
  {Beutler}}\ and\ \bibinfo {author} {\bibfnamefont {E.}~\bibnamefont
  {Di~Dio}},\ }\href {\doibase 10.1088/1475-7516/2020/07/048} {\bibfield
  {journal} {\bibinfo  {journal} {JCAP}\ }\textbf {\bibinfo {volume} {07}},\
  \bibinfo {pages} {048} (\bibinfo {year} {2020})},\ \Eprint
  {http://arxiv.org/abs/2004.08014} {arXiv:2004.08014 [astro-ph.CO]}
  \BibitemShut {NoStop}%
\bibitem [{\citenamefont {Prat}\ \emph {et~al.}(2018)\citenamefont {Prat} \emph
  {et~al.}}]{Prat:2017goa}%
  \BibitemOpen
  \bibfield  {author} {\bibinfo {author} {\bibfnamefont {J.}~\bibnamefont
  {Prat}} \emph {et~al.} (\bibinfo {collaboration} {DES}),\ }\href {\doibase
  10.1103/PhysRevD.98.042005} {\bibfield  {journal} {\bibinfo  {journal} {Phys.
  Rev. D}\ }\textbf {\bibinfo {volume} {98}},\ \bibinfo {pages} {042005}
  (\bibinfo {year} {2018})},\ \Eprint {http://arxiv.org/abs/1708.01537}
  {arXiv:1708.01537 [astro-ph.CO]} \BibitemShut {NoStop}%
\bibitem [{\citenamefont {Di~Dio}\ \emph {et~al.}(2013)\citenamefont {Di~Dio},
  \citenamefont {Montanari}, \citenamefont {Lesgourgues},\ and\ \citenamefont
  {Durrer}}]{DiDio:2013bqa}%
  \BibitemOpen
  \bibfield  {author} {\bibinfo {author} {\bibfnamefont {E.}~\bibnamefont
  {Di~Dio}}, \bibinfo {author} {\bibfnamefont {F.}~\bibnamefont {Montanari}},
  \bibinfo {author} {\bibfnamefont {J.}~\bibnamefont {Lesgourgues}}, \ and\
  \bibinfo {author} {\bibfnamefont {R.}~\bibnamefont {Durrer}},\ }\href
  {\doibase 10.1088/1475-7516/2013/11/044} {\bibfield  {journal} {\bibinfo
  {journal} {JCAP}\ }\textbf {\bibinfo {volume} {11}},\ \bibinfo {pages} {044}
  (\bibinfo {year} {2013})},\ \Eprint {http://arxiv.org/abs/1307.1459}
  {arXiv:1307.1459 [astro-ph.CO]} \BibitemShut {NoStop}%
\bibitem [{\citenamefont {Gaztanaga}\ \emph {et~al.}(2017)\citenamefont
  {Gaztanaga}, \citenamefont {Bonvin},\ and\ \citenamefont
  {Hui}}]{Gaztanaga:2015jrs}%
  \BibitemOpen
  \bibfield  {author} {\bibinfo {author} {\bibfnamefont {E.}~\bibnamefont
  {Gaztanaga}}, \bibinfo {author} {\bibfnamefont {C.}~\bibnamefont {Bonvin}}, \
  and\ \bibinfo {author} {\bibfnamefont {L.}~\bibnamefont {Hui}},\ }\href
  {\doibase 10.1088/1475-7516/2017/01/032} {\bibfield  {journal} {\bibinfo
  {journal} {JCAP}\ }\textbf {\bibinfo {volume} {01}},\ \bibinfo {pages} {032}
  (\bibinfo {year} {2017})},\ \Eprint {http://arxiv.org/abs/1512.03918}
  {arXiv:1512.03918 [astro-ph.CO]} \BibitemShut {NoStop}%
\bibitem [{\citenamefont {Bonvin}\ \emph {et~al.}(2016)\citenamefont {Bonvin},
  \citenamefont {Hui},\ and\ \citenamefont {Gaztanaga}}]{Bonvin:2015kuc}%
  \BibitemOpen
  \bibfield  {author} {\bibinfo {author} {\bibfnamefont {C.}~\bibnamefont
  {Bonvin}}, \bibinfo {author} {\bibfnamefont {L.}~\bibnamefont {Hui}}, \ and\
  \bibinfo {author} {\bibfnamefont {E.}~\bibnamefont {Gaztanaga}},\ }\href
  {\doibase 10.1088/1475-7516/2016/08/021} {\bibfield  {journal} {\bibinfo
  {journal} {JCAP}\ }\textbf {\bibinfo {volume} {08}},\ \bibinfo {pages} {021}
  (\bibinfo {year} {2016})},\ \Eprint {http://arxiv.org/abs/1512.03566}
  {arXiv:1512.03566 [astro-ph.CO]} \BibitemShut {NoStop}%
\bibitem [{\citenamefont {Hall}\ and\ \citenamefont
  {Bonvin}(2017)}]{Hall:2016bmm}%
  \BibitemOpen
  \bibfield  {author} {\bibinfo {author} {\bibfnamefont {A.}~\bibnamefont
  {Hall}}\ and\ \bibinfo {author} {\bibfnamefont {C.}~\bibnamefont {Bonvin}},\
  }\href {\doibase 10.1103/PhysRevD.95.043530} {\bibfield  {journal} {\bibinfo
  {journal} {Phys. Rev.}\ }\textbf {\bibinfo {volume} {D95}},\ \bibinfo {pages}
  {043530} (\bibinfo {year} {2017})},\ \Eprint
  {http://arxiv.org/abs/1609.09252} {arXiv:1609.09252 [astro-ph.CO]}
  \BibitemShut {NoStop}%
\bibitem [{\citenamefont {{Limber}}(1953)}]{1953ApJ...117..134L}%
  \BibitemOpen
  \bibfield  {author} {\bibinfo {author} {\bibfnamefont {D.~N.}\ \bibnamefont
  {{Limber}}},\ }\href {\doibase 10.1086/145672} {\bibfield  {journal}
  {\bibinfo  {journal} {\apj}\ }\textbf {\bibinfo {volume} {117}},\ \bibinfo
  {pages} {134} (\bibinfo {year} {1953})}\BibitemShut {NoStop}%
\bibitem [{\citenamefont {Kaiser}(1998)}]{Kaiser:1996tp}%
  \BibitemOpen
  \bibfield  {author} {\bibinfo {author} {\bibfnamefont {N.}~\bibnamefont
  {Kaiser}},\ }\href {\doibase 10.1086/305515} {\bibfield  {journal} {\bibinfo
  {journal} {Astrophys. J.}\ }\textbf {\bibinfo {volume} {498}},\ \bibinfo
  {pages} {26} (\bibinfo {year} {1998})},\ \Eprint
  {http://arxiv.org/abs/astro-ph/9610120} {arXiv:astro-ph/9610120} \BibitemShut
  {NoStop}%
\bibitem [{\citenamefont {Ghosh}\ \emph {et~al.}(2018)\citenamefont {Ghosh},
  \citenamefont {Durrer},\ and\ \citenamefont {Sellentin}}]{Ghosh:2018nsm}%
  \BibitemOpen
  \bibfield  {author} {\bibinfo {author} {\bibfnamefont {B.}~\bibnamefont
  {Ghosh}}, \bibinfo {author} {\bibfnamefont {R.}~\bibnamefont {Durrer}}, \
  and\ \bibinfo {author} {\bibfnamefont {E.}~\bibnamefont {Sellentin}},\ }\href
  {\doibase 10.1088/1475-7516/2018/06/008} {\bibfield  {journal} {\bibinfo
  {journal} {JCAP}\ }\textbf {\bibinfo {volume} {06}},\ \bibinfo {pages} {008}
  (\bibinfo {year} {2018})},\ \Eprint {http://arxiv.org/abs/1801.02518}
  {arXiv:1801.02518 [astro-ph.CO]} \BibitemShut {NoStop}%
\bibitem [{\citenamefont {Abbott}\ \emph {et~al.}(2018)\citenamefont {Abbott}
  \emph {et~al.}}]{DES:2017myr}%
  \BibitemOpen
  \bibfield  {author} {\bibinfo {author} {\bibfnamefont {T.~M.~C.}\
  \bibnamefont {Abbott}} \emph {et~al.} (\bibinfo {collaboration} {DES}),\
  }\href {\doibase 10.1103/PhysRevD.98.043526} {\bibfield  {journal} {\bibinfo
  {journal} {Phys. Rev. D}\ }\textbf {\bibinfo {volume} {98}},\ \bibinfo
  {pages} {043526} (\bibinfo {year} {2018})},\ \Eprint
  {http://arxiv.org/abs/1708.01530} {arXiv:1708.01530 [astro-ph.CO]}
  \BibitemShut {NoStop}%
\bibitem [{\citenamefont {Cai}\ and\ \citenamefont
  {Bernstein}(2012)}]{Cai_2012}%
  \BibitemOpen
  \bibfield  {author} {\bibinfo {author} {\bibfnamefont {Y.-C.}\ \bibnamefont
  {Cai}}\ and\ \bibinfo {author} {\bibfnamefont {G.}~\bibnamefont
  {Bernstein}},\ }\href {\doibase 10.1111/j.1365-2966.2012.20676.x} {\bibfield
  {journal} {\bibinfo  {journal} {Monthly Notices of the Royal Astronomical
  Society}\ }\textbf {\bibinfo {volume} {422}},\ \bibinfo {pages} {1045–1056}
  (\bibinfo {year} {2012})}\BibitemShut {NoStop}%
\bibitem [{\citenamefont {Blanchard}\ \emph {et~al.}(2020)\citenamefont
  {Blanchard} \emph {et~al.}}]{Euclid:2019clj}%
  \BibitemOpen
  \bibfield  {author} {\bibinfo {author} {\bibfnamefont {A.}~\bibnamefont
  {Blanchard}} \emph {et~al.} (\bibinfo {collaboration} {Euclid}),\ }\href
  {\doibase 10.1051/0004-6361/202038071} {\bibfield  {journal} {\bibinfo
  {journal} {Astron. Astrophys.}\ }\textbf {\bibinfo {volume} {642}},\ \bibinfo
  {pages} {A191} (\bibinfo {year} {2020})},\ \Eprint
  {http://arxiv.org/abs/1910.09273} {arXiv:1910.09273 [astro-ph.CO]}
  \BibitemShut {NoStop}%
\bibitem [{\citenamefont {Aghamousa}\ \emph {et~al.}(2016)\citenamefont
  {Aghamousa} \emph {et~al.}}]{DESI:2016fyo}%
  \BibitemOpen
  \bibfield  {author} {\bibinfo {author} {\bibfnamefont {A.}~\bibnamefont
  {Aghamousa}} \emph {et~al.} (\bibinfo {collaboration} {DESI}),\ }\href@noop
  {} {\  (\bibinfo {year} {2016})},\ \Eprint {http://arxiv.org/abs/1611.00036}
  {arXiv:1611.00036 [astro-ph.IM]} \BibitemShut {NoStop}%
\bibitem [{\citenamefont {Dey}\ \emph {et~al.}(2019)\citenamefont {Dey},
  \citenamefont {Schlegel}, \citenamefont {Lang}, \citenamefont {Blum},
  \citenamefont {Burleigh}, \citenamefont {Fan}, \citenamefont {Findlay},
  \citenamefont {Finkbeiner}, \citenamefont {Herrera}, \citenamefont {Juneau},\
  and\ \citenamefont {et~al.}}]{Dey_2019}%
  \BibitemOpen
  \bibfield  {author} {\bibinfo {author} {\bibfnamefont {A.}~\bibnamefont
  {Dey}}, \bibinfo {author} {\bibfnamefont {D.~J.}\ \bibnamefont {Schlegel}},
  \bibinfo {author} {\bibfnamefont {D.}~\bibnamefont {Lang}}, \bibinfo {author}
  {\bibfnamefont {R.}~\bibnamefont {Blum}}, \bibinfo {author} {\bibfnamefont
  {K.}~\bibnamefont {Burleigh}}, \bibinfo {author} {\bibfnamefont
  {X.}~\bibnamefont {Fan}}, \bibinfo {author} {\bibfnamefont {J.~R.}\
  \bibnamefont {Findlay}}, \bibinfo {author} {\bibfnamefont {D.}~\bibnamefont
  {Finkbeiner}}, \bibinfo {author} {\bibfnamefont {D.}~\bibnamefont {Herrera}},
  \bibinfo {author} {\bibfnamefont {S.}~\bibnamefont {Juneau}}, \ and\ \bibinfo
  {author} {\bibnamefont {et~al.}},\ }\href {\doibase 10.3847/1538-3881/ab089d}
  {\bibfield  {journal} {\bibinfo  {journal} {The Astronomical Journal}\
  }\textbf {\bibinfo {volume} {157}},\ \bibinfo {pages} {168} (\bibinfo {year}
  {2019})}\BibitemShut {NoStop}%
\bibitem [{\citenamefont {Peebles}(2002)}]{Peebles:2002iq}%
  \BibitemOpen
  \bibfield  {author} {\bibinfo {author} {\bibfnamefont {P.~J.~E.}\
  \bibnamefont {Peebles}},\ }in\ \href@noop {} {\emph {\bibinfo {booktitle}
  {{37th Rencontres de Moriond on the Cosmological Model}}}}\ (\bibinfo {year}
  {2002})\ \Eprint {http://arxiv.org/abs/astro-ph/0208037}
  {arXiv:astro-ph/0208037} \BibitemShut {NoStop}%
\bibitem [{\citenamefont {Pullen}\ \emph {et~al.}(2016)\citenamefont {Pullen},
  \citenamefont {Alam}, \citenamefont {He},\ and\ \citenamefont
  {Ho}}]{Pullen:2015vtb}%
  \BibitemOpen
  \bibfield  {author} {\bibinfo {author} {\bibfnamefont {A.~R.}\ \bibnamefont
  {Pullen}}, \bibinfo {author} {\bibfnamefont {S.}~\bibnamefont {Alam}},
  \bibinfo {author} {\bibfnamefont {S.}~\bibnamefont {He}}, \ and\ \bibinfo
  {author} {\bibfnamefont {S.}~\bibnamefont {Ho}},\ }\href {\doibase
  10.1093/mnras/stw1249} {\bibfield  {journal} {\bibinfo  {journal} {Mon. Not.
  Roy. Astron. Soc.}\ }\textbf {\bibinfo {volume} {460}},\ \bibinfo {pages}
  {4098} (\bibinfo {year} {2016})},\ \Eprint {http://arxiv.org/abs/1511.04457}
  {arXiv:1511.04457 [astro-ph.CO]} \BibitemShut {NoStop}%
\end{thebibliography}%

\end{document}